\documentclass[twoside,twocolumn,9pt]{article}
\usepackage{extsizes}
\usepackage[super,sort&compress,comma]{natbib} 
\usepackage[version=3]{mhchem}
\usepackage[left=1.5cm, right=1.5cm, top=1.785cm, bottom=2.0cm]{geometry}
\usepackage{balance}
\usepackage{mathptmx}
\usepackage{sectsty}
\usepackage{graphicx} 
\usepackage{lastpage}
\usepackage[format=plain,justification=justified,singlelinecheck=false,font={stretch=1.125,small,sf},labelfont=bf,labelsep=space]{caption}
\usepackage{float}
\usepackage{fancyhdr}
\usepackage{fnpos}
\usepackage[english]{babel}
\addto{\captionsenglish}{%
  
}
\usepackage{array}
\usepackage{droidsans}
\usepackage{charter}
\usepackage[T1]{fontenc}
\usepackage[usenames,dvipsnames]{xcolor}
\usepackage{setspace}
\usepackage[compact]{titlesec}
\usepackage{hyperref}
\usepackage{epstopdf}%This line makes .eps figures into .pdf - please comment out if not required.

\definecolor{cream}{RGB}{222,217,201}

\begin{document}

\pagestyle{fancy}
\thispagestyle{plain}
\fancypagestyle{plain}{
%%%HEADER%%%
\renewcommand{\headrulewidth}{0pt}
}
%%%END OF HEADER%%%

%%%PAGE SETUP - Please do not change any commands within this section%%%
\makeFNbottom
\makeatletter
\renewcommand\LARGE{\@setfontsize\LARGE{15pt}{17}}
\renewcommand\Large{\@setfontsize\Large{12pt}{14}}
\renewcommand\large{\@setfontsize\large{10pt}{12}}
\renewcommand\footnotesize{\@setfontsize\footnotesize{7pt}{10}}
\makeatother

\renewcommand{\thefootnote}{\fnsymbol{footnote}}
\renewcommand\footnoterule{\vspace*{1pt}% 
\color{cream}\hrule width 3.5in height 0.4pt \color{black}\vspace*{5pt}} 
\setcounter{secnumdepth}{5}

\makeatletter 
\renewcommand\@biblabel[1]{#1}            
\renewcommand\@makefntext[1]% 
{\noindent\makebox[0pt][r]{\@thefnmark\,}#1}
\makeatother 
\renewcommand{\figurename}{\small{Fig.}~}
\sectionfont{\sffamily\Large}
\subsectionfont{\normalsize}
\subsubsectionfont{\bf}
\setstretch{1.125} %In particular, please do not alter this line.
\setlength{\skip\footins}{0.8cm}
\setlength{\footnotesep}{0.25cm}
\setlength{\jot}{10pt}
\titlespacing*{\section}{0pt}{4pt}{4pt}
\titlespacing*{\subsection}{0pt}{15pt}{1pt}
%%%END OF PAGE SETUP%%%

%%%FOOTER%%%
\fancyfoot{}
\fancyfoot[LO,RE]{\vspace{-7.1pt}\includegraphics[height=9pt]{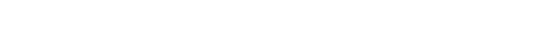}}
\fancyfoot[CO]{\vspace{-7.1pt}\hspace{13.2cm}\includegraphics{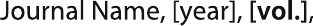}}
\fancyfoot[CE]{\vspace{-7.2pt}\hspace{-14.2cm}\includegraphics{head_foot/RF}}
\fancyfoot[RO]{\footnotesize{\sffamily{1--\pageref{LastPage} ~\textbar  \hspace{2pt}\thepage}}}
\fancyfoot[LE]{\footnotesize{\sffamily{\thepage~\textbar\hspace{3.45cm} 1--\pageref{LastPage}}}}
\fancyhead{}
\renewcommand{\headrulewidth}{0pt} 
\renewcommand{\footrulewidth}{0pt}
\setlength{\arrayrulewidth}{1pt}
\setlength{\columnsep}{6.5mm}
\setlength\bibsep{1pt}
%%%END OF FOOTER%%%

%%%FIGURE SETUP - please do not change any commands within this section%%%
\makeatletter 
\newlength{\figrulesep} 
\setlength{\figrulesep}{0.5\textfloatsep} 

\newcommand{\topfigrule}{\vspace*{-1pt}% 
\noindent{\color{cream}\rule[-\figrulesep]{\columnwidth}{1.5pt}} }

\newcommand{\botfigrule}{\vspace*{-2pt}% 
\noindent{\color{cream}\rule[\figrulesep]{\columnwidth}{1.5pt}} }

\newcommand{\dblfigrule}{\vspace*{-1pt}% 
\noindent{\color{cream}\rule[-\figrulesep]{\textwidth}{1.5pt}} }

\makeatother
%%%END OF FIGURE SETUP%%%

%%%TITLE, AUTHORS AND ABSTRACT%%%
\twocolumn[
  \begin{@twocolumnfalse}
{\includegraphics[height=30pt]{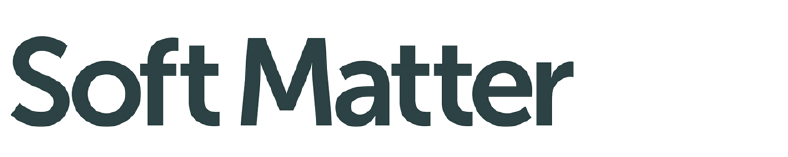}\hfill\raisebox{0pt}[0pt][0pt]{\includegraphics[height=55pt]{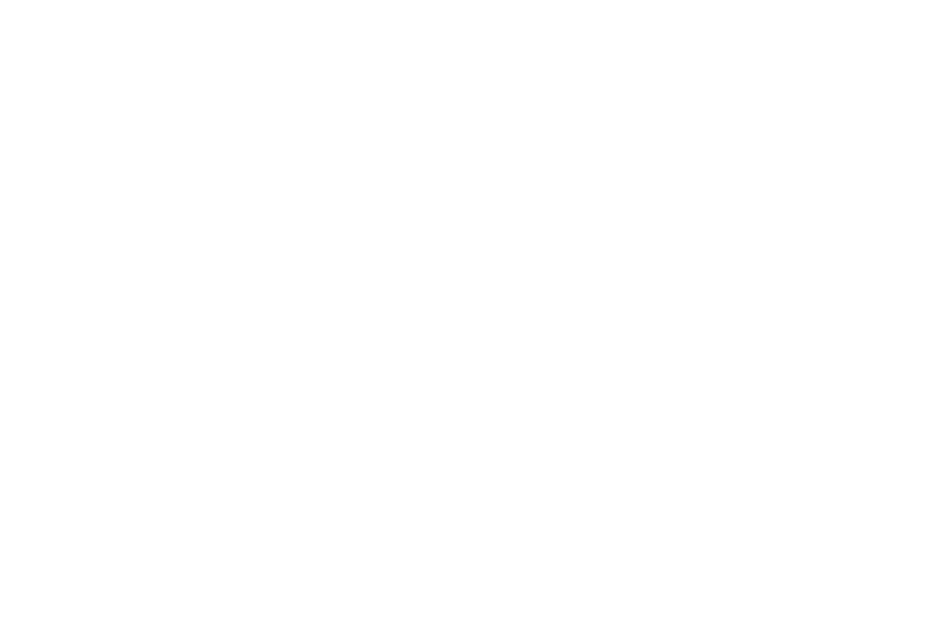}}\\[1ex]
\includegraphics[width=18.5cm]{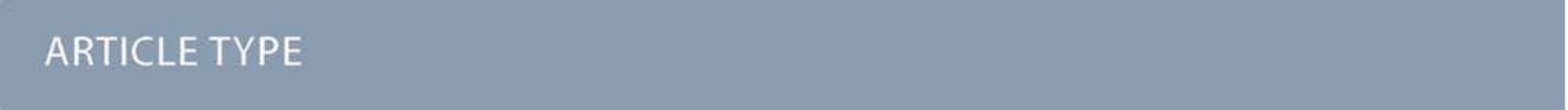}}\par
\vspace{1em}
\sffamily
\begin{tabular}{m{4.5cm} p{13.5cm} }

\includegraphics{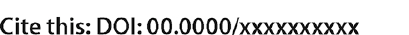} & \noindent\LARGE{\textbf{Enhanced clamshell swimming with asymmetric beating at low Reynolds number}} \\%Article title goes here instead of the text "This is the title"
\vspace{0.3cm} & \vspace{0.3cm} \\

 & \noindent\large{Shiyuan Hu,$^{abc}$ Jun Zhang,$^{abc}$ and Michael J. Shelley{$^{\ast}$}{$^{ad}$}} \\%Author names go here instead of "Full name", etc.

\includegraphics{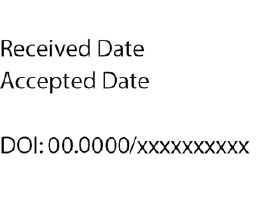} & \noindent\normalsize{A single flexible filament can be actuated to escape from the scallop theorem and generate net propulsion at low Reynolds number. In this work, we study the dynamics of a simple boundary-driven multi-filament swimmer, a two-arm clamshell actuated at the hinged point, using a nonlocal slender body approximation with full hydrodynamic interactions. We first consider an elastic clamshell consisted of flexible filaments with intrinsic curvature, and then build segmental models consisted of rigid segments connected by different mechanical joints with different forms of response torques. The simplicity of the system allows us to fully explore the effect of various parameters on the swimming performance. Optimal included angles and elastoviscous numbers are identified. The segmental models capture the characteristic dynamics of the elastic clamshell. We further demonstrate how the swimming performance can be significantly enhanced by the asymmetric beating patterns induced by biased torques.} 

\end{tabular}

\end{@twocolumnfalse} \vspace{0.6cm}
]
%%%END OF TITLE, AUTHORS AND ABSTRACT%%%

%%%FONT SETUP - please do not change any commands within this section
\renewcommand*\rmdefault{bch}\normalfont\upshape
\rmfamily
\section*{}
\vspace{-1cm}

%%%FOOTNOTES%%%

\footnotetext{\textit{$^{a}$~Applied Mathematics Lab, Courant Institute of Mathematical Sciences, New York University, New York, NY 10012, USA.}}
\footnotetext{\textit{$^{b}$~Department of Physics, New York University, New York, NY 10003, USA.}}
\footnotetext{\textit{$^{c}$~NYU-ECNU Institute of Physics at NYU Shanghai, Shanghai 200062, China.}}
\footnotetext{\textit{$^{d}$~Center for Computational Biology, Flatiron Institute, New York, NY 10010, USA. E-mail: mshelley@flatironinstitute.org}}

%Please use \dag to cite the ESI in the main text of the article.
%If you article does not have ESI please remove the the \dag symbol from the title and the footnotetext below.
\footnotetext{\dag~Electronic Supplementary Information (ESI) available: [details of any supplementary information available should be included here]. See DOI: 10.1039/cXsm00000x/}
%additional addresses can be cited as above using the lower-case letters, c, d, e... If all authors are from the same address, no letter is required

%%%END OF FOOTNOTES%%%

%%%MAIN TEXT%%%%
%%%%%%%%%%%%%%%%%%%%%%%%%%%%%%%%%%
\section{Introduction}\label{sec.1}
%%%%%%%%%%%%%%%%%%%%%%%%%%%%%%%%%%
Reciprocal motions at low Reynolds number (Re) in Newtonian fluids cannot generate net translations, a fact known as the scallop theorem~\cite{Purcell77,Lauga11}. Non-reciprocal kinematics that breaks time reversal symmetry can lead to locomotion. In the biological realm many microorganisms use elastic appendages to swim. The bacteria flagella are helical shaped and driven by rotary motors at the base~\cite{Berg73}. Eukaryotic flagella and cilia are internally actuated by distributed molecular motors and can display various beating patterns~\cite{Alberts15,Han18,Chakrabarti19}. For example, spermatozoa generate wavelike deformations along their flagella~\cite{Brennen77}. For a ciliated microorganism, each cilium beats asymmetrically: the cilium extends during the power stoke pushing the fluids and bends with larger deformation, thus reducing the drag during the recovery stroke~\cite{Blake74}. The biflagellate alga \textit{Chlamydomonas} adapts an effective gait during locomotion that resembles `breaststroke' swimming as its two flagella asymmetrically bent during the power and recovery strokes~\cite{Mitchell00,Goldstein09}.

The design and optimization of artificial swimmers are important research areas, related to biological locomotion~\cite{Childress81}, with applications to pumping, mixing, and cargo delivery at low Re~\cite{Kim04,Darnton04,Weibel05,Ebbens10,Ceylan17}. Biologically inspired microswimmers with synthetic appendages have been realized and tested experimentally~\cite{Dreyfus05,Williams14,Maier16,Ali17,Huang19}. Simple swimmers using discrete degrees of freedom to generate non-reciprocal motions have also been studied, as demonstrated by Purcell’s three-link swimmer~\cite{Purcell77,Becker03} and the three-sphere swimmer~\cite{Najafi04,Leoni09}. For the optimization of swimming and propulsion performance, earlier works include finding the optimal waveform for flagellum~\cite{Lighthill75,Pironneau74} and the optimal geometry of the swimming cell~\cite{Higdon79,Higdon2_79}. More recent studies on swimming optimization have been devoted to the stroke patterns of the three-link swimmer~\cite{Tam07}, beating patterns of cilia~\cite{Osterman11}, and swimming gaits of \textit{Chlamydomonas}~\cite{Tam11}.

A simple design strategy of artificial swimmers involves elastic filaments with boundary actuations, such as angular or positional oscillations at the filament's ends, that send travelling waves along the filaments~\cite{Wiggins98,Yu06,Lauga07}. The effects of various mechanisms on swimming performance have been studied, such as the hydrodynamic interactions~\cite{Singh18,Elfasi18,Liu20}, the number of filaments~\cite{Ye13}, and the filament intrinsic curvature~\cite{Ye13,Liu20}. It has been demonstrated in experiments that the velocity of a swimmer propelled by multiple filaments may be enhanced by intrinsically curved filaments~\cite{Ye13}, which was subsequently explored in numerical simulations based on discrete elastic rod model~\cite{Liu20}. The enhanced swimming was attributed to the alignment of the propulsion directions of the filaments and their tilt angles. An elastic clamshell moving in two-dimensional Stokesian fluid has been constructed using a bead-spring model and found to translate from the hinge point to the open side~\cite{Choudhary18}. However, the mechanical design and optimization of low-Re swimmers with multiple filaments remains largely unexplored. 

In this paper, we study the dynamics of a simple multi-filament swimmer, a clamshell consisting of two arms hinged at one common end without load, moving in a three-dimensional Stokesian fluid. The slender and inextensible filaments are modeled using a non-local slender body approximation with full hydrodynamic interactions (HIs)~\cite{Johnson80}. We also construct a segmental model with finite degrees of freedom by replacing the flexible filaments with jointed rigid segments. As functions of the relative deflection angle between the rigid segments, different forms of passive response torques at the joints are considered, which add rotational resistance to the filament dynamics. In particular, we consider biased response torques that mimic flexible filaments with nonzero intrinsic curvature. The asymmetric beating patterns generated by the biased torques significantly increase the swimming speed and efficiency. 

We present the theoretical formulation of the elastic clamshell in Sec.~\ref{sec.2.1} and that of segmental model in Sec.~\ref{sec.2.2}. The derivation of boundary conditions and details on numerical methods are included in Appx.~\ref{sec.a} and \ref{sec.b}. We discuss our main results in Sec.~\ref{sec.3} and finally conclude this work with remarks in Sec.~\ref{sec.4}. 

%%%%%%%%%%%%%%%%%%%%%%%%%%%%%%%%%%%%%%%%%%%%%%%%%%%%%%%%%%%
\section{Theoretical formulation} \label{sec.2}
%%%%%%%%%%%%%%%%%%%%%%%%%%%%%%%%%%%%%%%%%%%%%%%%%%%%%%%%%%%
\subsection{Elastic clamshell}\label{sec.2.1}
%%%%%%%%%%%%%%%%%%%%%%%%%%%%%%%%%%%%%%%%%%%%%%%%%%%%%%%%%%%
\begin{figure}[t]
\centering
\includegraphics[bb=0 5 260 130, scale=0.8, draft=false]{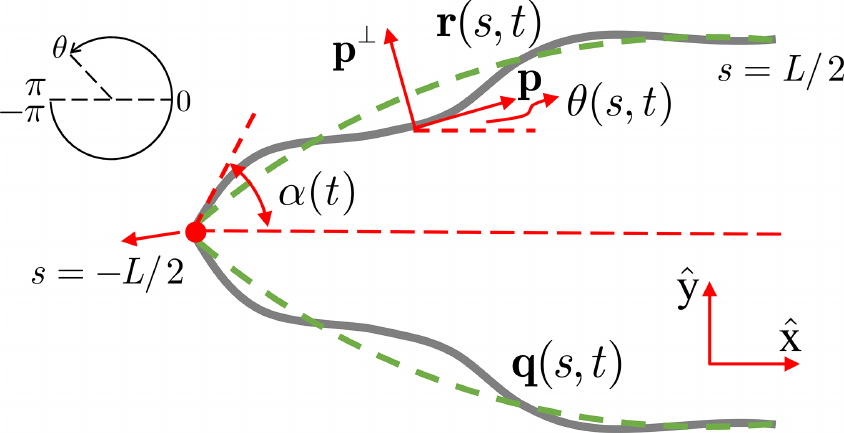}
\caption{Schematic of an elastic clamshell swimmer. Two flexible filaments with intrinsic curvature are hinged at $s=-L/2$ and mirror-symmetric about $x$ axis. The swimmer is driven by a time-varying angle $\alpha(t)$ at $s=-L/2$ and translating along $x$ axis. The green dashed curves indicate the shapes of the filaments at rest. The tangent angle $\theta$ is defined to increase counterclockwise from $-\pi$ to $\pi$.}
\label{fig1}
\end{figure}
%%%%%%%%%%%%%%%%%
Consider a slender, inextensible and elastic filament of radius $a$, length $L$ (with aspect ratio $\epsilon = a/L \ll 1$), and bending rigidity $B$, moving in a quiescent three-dimensional Stokesian fluid of viscosity $\mu$ with the filament's motion confined to a two-dimensional plane. The filaments have an intrinsic curvature $\kappa_0$, taken as constant along the filament. Denote the filament centerline by $\textbf{r}(s)$ with the signed arc length $s \in [-L/2, L/2]$. The unit tangent vector $\textbf{p} = \textbf{r}_s = \cos\theta\hat{\textbf{x}}+\sin\theta\hat{\textbf{y}}$ with $\theta$ the tangent angle. The unit normal vector $\textbf{p}^{\perp}=\textbf{p}_s/\theta_s=-\sin\theta\hat{\textbf{x}}+\cos\theta\hat{\textbf{y}}$. We describe the filament as an Euler-Bernoulli beam with its energy given by
%%%%
\begin{equation}\label{eq2.1}
\mathcal{E}=\frac{1}{2}B\int_{-L/2}^{L/2}(\kappa-\kappa_0)^{2}ds+\frac{1}{2}\int_{-L/2}^{L/2}T(|\textbf{r}_s|^2-1)ds,
\end{equation}
%%%%
where the local curvature $\kappa=\theta_s$. The first term is the bending energy and the second term imposes the inextensibility of the filament with $T$ the filament tension. The filament force per unit length \textbf{f} upon the fluid can be derived from the variation of $\mathcal{E}$ with respect to a small and arbitrary shape deformation $\delta \textbf{r}$, i.e., $\delta \mathcal{E} = -\int_{-L/2}^{L/2} \textbf{f}\cdot \delta \textbf{r}\,ds$, leading to,
\begin{equation}\label{eq2.2}
\textbf{f} =-B\left[\textbf{r}_{ssss}+\kappa_0\left(\kappa\textbf{p}\right)_{s}\right]+\left(T\textbf{p}\right)_{s},
\end{equation}
From non-local slender body theory~\cite{Johnson80}, the velocity of the filament centerline $\textbf{r}_{t}$ is governed by a balance of filament forces and viscous drag:
\begin{equation}\label{eq2.3}
8 \pi \mu (\textbf{r}_{t}-\textbf{U})=\left[c\left(\textbf{I}+\textbf{p}\textbf{p}\right)+2(\textbf{I}-\textbf{p}\textbf{p})\right]\textbf{f},
\end{equation}
where $c = |\ln(\epsilon^{2}e)|$ and $\textbf{U}$ is the nonlocal flow field induced by the filaments. Using the inextensibility condition, $\textbf{r}_{s} \cdot \textbf{r}_{st}=0$, Eq.~(\ref{eq2.3}) can be manipulated to give the equation for the tension, 
\begin{equation}\label{eq2.4}
\begin{aligned}
2cT_{ss}-(c+2)\theta_{s}^{2}T =&-8\pi\mu\textbf{U}_s\cdot \textbf{p}-6cB\theta_{ss}^{2}-(7c+2)B\theta_{s}\theta_{sss} \\
&+(c+2)B\theta_{s}^{4}-(c+2)B\kappa_0\theta_{s}^{3}+2\kappa_0 B\theta_{sss},
\end{aligned}
\end{equation}
The evolution equation of the tangent angle $\theta$ can be derived from $\theta_{t} = \textbf{r}_{st}\cdot \textbf{p}^{\perp}$,
\begin{equation}\label{eq2.5}
\begin{aligned}
8\pi\mu\theta_{t}+(c+2)B\theta_{ssss} = 8\pi\mu \textbf{U}_{s} \cdot \textbf{p}^{\perp} + (9c+6)B\theta_{s}^{2}\theta_{ss} \\
+(3c+2)T_{s}\theta_{s}+(c+2)T\theta_{ss}-(4c+4)\kappa_0 B\theta_{s}\theta_{ss}.
\end{aligned}
\end{equation}
Equations~(\ref{eq2.3}), (\ref{eq2.4}), and (\ref{eq2.5}) are the governing equations of the dynamics of a flexible filament with intrinsic curvature in Stokesian flow.  

We construct a clamshell swimmer with two mirror-symmetric flexible filaments jointed at $s=-L/2$ (Fig.~\ref{fig1}). The swimmer is driven by a sinusoidally-oscillating angle at the hinged point between the filaments:
%%%%
\begin{equation}\label{eq2.6}
\alpha(t) = \alpha_{0}\left[\sin\left(2\pi t/\tau_0\right)+1\right],
\end{equation}
%%%%
where $\tau_0$ is the oscillation period, and $\alpha_0$ is the actuation amplitude and $\alpha \in[0, 2\alpha_0]$. Here, $\alpha_0$ is limited to avoid filament intersections. Due to the mirror symmetry, we only consider the dynamics of the upper filament $\textbf{r}(s,t)$. The background velocity $\textbf{U}$ in Eq.~(\ref{eq2.3}) is the flow induced by the motion of the two filaments, which are associated with distributions of fundamental solutions of Stokes equation along the filament centerline and which capture nonlocal hydrodynamic interactions. There are two contributions to $\textbf{U}$, $\textbf{U}(s) = \textbf{V}_1[\textbf{f}](s)+\textbf{V}_2[\textbf{f}](s)$. The flow field induced by the filament upon itself, $\textbf{V}_1$, is given by
%%%%
\begin{equation}\label{eq2.7}
\textbf{V}_1(s) = \frac{1}{8\pi\mu}\int_{-L/2}^{L/2} \left[\frac{\textbf{I}+\hat{\textbf{R}}\hat{\textbf{R}}}{|\textbf{R}|}\textbf{f}(s')-\frac{\textbf{I}+\textbf{p}\textbf{p}}{|s-s'|}\textbf{f}(s)\right]\,ds',
\end{equation}
%%%%
where $\textbf{R} = \textbf{r}(s)-\textbf{r}(s')$. The flow field induced by the other filament is given by
%%%%%
\begin{equation}\label{eq2.8}
\textbf{V}_2(s) = \frac{1}{8\pi\mu}\int_{-L/2}^{L/2}\frac{\textbf{I}+\hat{\textbf{R}}\hat{\textbf{R}}}{|\textbf{R}|}\textbf{f}(s')\,ds',
\end{equation}
%%%%%
where $\textbf{R} = \textbf{r}(s)-\textbf{q}(s')$ and $\textbf{q}$ is the position of the other filament. The system can be non-dimensionalized using length $L$, force $B/L^{2}$, and time $\tau_0$. One resulting dimensionless parameter is the elastoviscous number, $\eta = L\Big/\left(\frac{B \tau_0}{8\pi \mu}\right)^{1/4}$. In the small $\eta$ regime, the filaments are nearly rigid with elastic force dominating viscous force; in the large $\eta$ regime, the filaments are very flexible with viscous stress dominating elastic stress. 

The necessary constraints and resulting boundary conditions are given in Appendix~\ref{sec.a}. We solve the set of partial differential equations given by Eqs.~(\ref{eq2.3})--(\ref{eq2.5}) numerically based on a second-order finite difference scheme~\cite{Tornberg04}. Due to the nonlinearity we use Newton's method to solve the tension equation. To avoid the stability limit for the time-step size arising from the fourth-order derivative, we use a second-order implicit/explicit backward differentiation scheme for the time stepping and treat the fourth-order derivative implicitly. More details of the numerical methods are given in Appendix~\ref{sec.b}. The control parameters include the oscillating amplitude $\alpha_0$, elastoviscous number $\eta$, and the intrinsic curvature $\kappa_0$.  

%%%%%%%%%%%%%%%%%%
\subsection{Segmental model with rigid filaments}\label{sec.2.2}
\begin{figure}[b]
\centering
\includegraphics[bb=0 0 250 80, scale=0.97,draft=false]{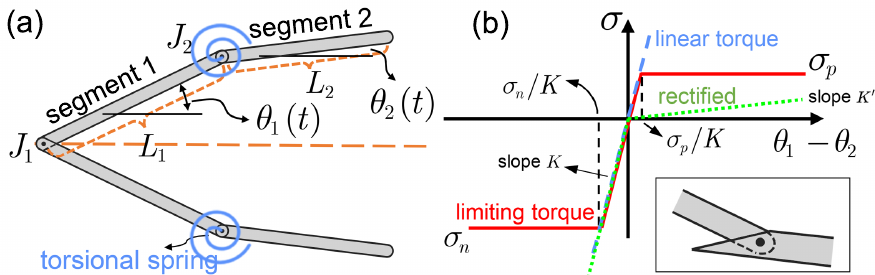}
\caption{Segmental model of the elastic clamshell. (a) Rigid filaments are connected using mechanical links such as torsional springs at $J_2$. (b) The linear torque (blue dashed line), limiting torque (red line), and rectified torque (green dotted line) applied at $J_2$ as functions of the deflected angle $\theta_1-\theta_2$. The inset shows a possible mechanical hinge that cannot open outwards with a constraint torque.}
\label{fig2}
\end{figure}

%%%%%%%%%%%%%%%%%%
We develop a second and different model by replacing each flexible filament with two rigid segments of different lengths (see schematic in Fig.~\ref{fig2}a). Below we use subscript integer to denote quantities associated with segment 1 and 2. The total length of the two segments is fixed, $L_1 + L_2 = L$, and we vary their length ratio, $\gamma = L_1/L$. The two segments are connected with different mechanical joints at $J_2$. The orientation of segment 1 is kinematically driven with the angle dynamics the same as Eq.~(\ref{eq2.6}) and segment 2 is passively responding (subjected to the rotational resistance applied by the mechanical joint). The centerline of each segment is described by a straight line $\textbf{r}_{k} = \textbf{r}_{k}^c+s_{k}\textbf{p}_{k}$ for $k=$1, 2, where $\textbf{r}_{k}^c$ is the center-of-mass (COM) position and $\textbf{p}_{k} = (\cos\theta_{k}, \sin\theta_{k})$ with $\theta_{k}$ the segment orientation.  
%%%
%%%

The dynamics of each rigid segment is governed by Eq.~(\ref{eq2.3}) with the constraint that the total $x$-component force is zero, 
%%%
\begin{equation}\label{eq2.9}
\int_{-L_1/2}^{L_1/2}f_1^x\,ds_1 + \int_{-L_2/2}^{L_2/2}f_2^x\,ds_2 = 0.
\end{equation}
%%%
The COM velocity of segment 1 can be decomposed into two components: the translation with $J_1$ and the COM rotation around $J_1$. Since $J_1$ only moves along the $x$-axis due to symmetry, the $y$-component COM velocity of segment 1 is determined by the rotation around $J_1$,
%%%
\begin{equation}\label{eq2.10}
\dot{y}_1^c = \frac{L_1}{2} \dot{\theta}_1 (t) \cos\theta_1,
\end{equation}
%%%
where $\theta_1$ is prescribed by Eq.~(\ref{eq2.6}), $\theta_1 = \alpha(t)$. The dynamics is further subjected to the constraints that the velocities of the two segments at $J_2$ are the same,
%%%
\begin{equation}\label{eq2.11}
\dot{\textbf{r}}_1^c+\frac{L_1}{2}\dot{\textbf{p}}_1 = \dot{\textbf{r}}_2^c-\frac{L_2}{2}\dot{\textbf{p}}_2.
\end{equation}
%%%
Finally, we balance the hydrodynamic torque acting upon segment 2 with the response torques by the mechanical joints at $J_2$,
%%%
\begin{equation}\label{eq2.12}
\int_{-L_2/2}^{L_2/2}(s_2+L_2/2)\textbf{p}_2\times \textbf{f}_2\,ds_2 = \sigma[\theta_1, \theta_2]\hat{\textbf{z}}.
\end{equation}
%%%
The simplest $\sigma$ is a linear function of the relative angular deflection (blue dashed line in Fig.~\ref{fig2}b): $\sigma = K\Delta \theta$, where $\Delta \theta = \theta_1-\theta_2$. The linear torque can be generated by a torsional spring with elastic modulus $K$. Another variant is a limiting torque implemented as a piecewise function (red line in Fig.~\ref{fig2}b): 
%%%
\begin{equation}\label{eq_limiting_torque}
\sigma = 
\begin{cases}
\sigma_n, & \Delta\theta < \sigma_n/K; \\
K\Delta\theta, & \sigma_n/K \le \Delta\theta \le \sigma_p/K;\\
\sigma_p, & \Delta\theta > \sigma_p/K.
\end{cases}
\end{equation}
%%%
The above torque may be generated by a mechanical joint similar to the one in a `torque wrench'. We also consider a rectified torque (green dotted line), which has a different elastic modulus $K'$ when $\Delta\theta > 0$ and $K' \ll K$. These different forms of torques do not add a hard constraint on $\Delta \theta$. To implement a locked hinge similar to the one shown in Fig.~\ref{fig2} inset, a constraint torque $\sigma_c$ is needed to prevent $\theta_2$ from increasing further when $\Delta \theta$ becomes smaller than a threshold $\theta_m$ during the power stroke, i.e., $\sigma = \sigma_c$ when $\Delta \theta < \theta_m$ and $\dot{\theta}_1<0$; $\sigma = K\Delta\theta$ otherwise. The constraint torque is exactly the hydrodynamic torque needed to keep $\dot{\theta}_2 = \dot{\theta}_1$ ($\Delta \theta$ fixed), which can be written out explicitly if the nonlocal integral in Eq.~(\ref{eq2.3}) is ignored,
%%%
\begin{equation}
\sigma_c = \frac{\pi\mu L_2^2}{c+2}\left[4(-\dot{x}_2^c\sin\theta_2+\dot{y}_2^c\cos\theta_2) + \frac{2}{3}L_2 \dot{\theta}_1\right].
\end{equation}
%%%
Compared with the infinite-dimensional elastic clamshell, the segmental model has only $4$ discrete degrees of freedom, described by $\textbf{r}_1^c$, $\theta_1$, $\theta_2$.

We non-dimensionlize the system by scaling lengths on $L$, time on $\tau_0$, and forces on $\mu L^2\tau_0^{-1}$. The dimensionless control parameters include the segment length ratio $\gamma$, the elastoviscous number, $\eta = L\big/(K\tau_0/\mu)^{1/3}$, and the torque-bias parameters $\sigma_n$, $\sigma_p$, $K/K'$, and $\theta_m$. At each time step, a linear system from Eq.~(\ref{eq2.3}), (\ref{eq2.9})--(\ref{eq2.12}) is solved to determine $\dot{\textbf{r}}_k$, $\dot{\theta}_k$, and $\textbf{f}_k$. The set of ordinary differential equations thus obtained are then evolved using a 4th-order Runge-Kutta scheme. For both the elastic clamshell and the segmental model, the filaments are not self-intersecting for the range of parameters explored in this work. 

%%%%%%%%%%%%%%%%%%
\section{Results and Discussion}\label{sec.3}
%%%%%%%%%%%%%%%%%%
\subsection{Elastic clamshell}
%%%%%%%%%%%%%%%%%%

%%%%%%
\begin{figure}[t]
\centering
\includegraphics[bb=0 10 245 134, scale=1.0,draft=false]{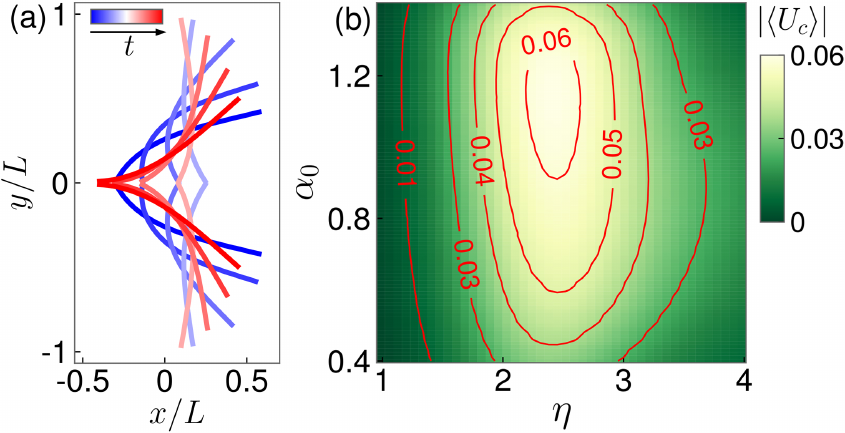}
\caption{(a) Time-lapse of the deformations of the elastic clamshell over one oscillation period with $\eta = 3$, $\kappa_0=0$, $\alpha_0 = 1.1$, and $c=15.0$ during the recovery stroke (blue) and power stroke (red). Time runs from blue to red. See supplemental videos showing motions of the elastic clamshell with different parameters. (b) Swimming speed $|\langle U_c \rangle|$ shown as a single-peaked function of $\alpha_0$ and $\eta$.}
\label{fig3}
\end{figure}
%%%%%
The motion of the clamshell swimmer in each period consists of a recovery stroke with the two filaments opening ($\dot{\alpha}>0$) and a power stroke with the two filaments closing ($\dot{\alpha}<0$). The filaments are bent inward due to the viscous drag during the recovery stroke and the COM moves towards the $+x$ direction; during the power stroke, the filaments are bent outward and the COM moves towards the $-x$ direction. This asymmetry in the filament's deformation leads to a net translation after one period (Fig.~\ref{fig3}a). We compute the time-averaged COM velocity $\langle \textbf{U}_c\rangle = \langle U_c\rangle \hat{\textbf{x}}$, where the time-averaged swimming speed is given by 
%%%%%%
\begin{figure}[b]
\centering
\includegraphics[bb=0 5 242 88, scale=1.03,draft=false]{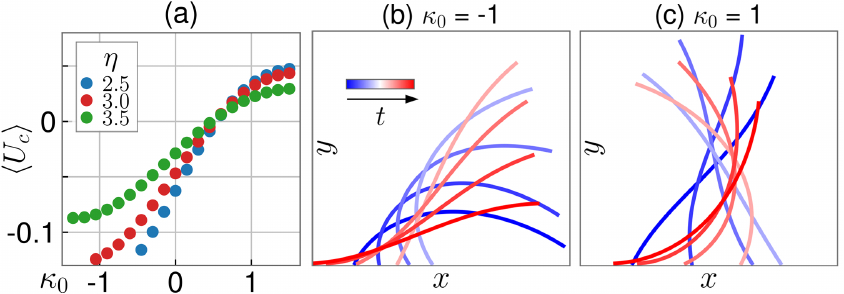}
\caption{(a) $\langle U_c \rangle$ as a function of $\kappa_0$ for three different values of $\eta$. (b), (c) Time lapse of the deformations of the upper filament for (b) $\kappa_0 = -1$ and (c) $\kappa_0 = 1$ with $\eta=3.0$. Blue: recovery stroke; red: power stroke.}
\label{fig4} 
\end{figure}
%%%%%
%%%
\begin{equation}\label{eq3.1}
\langle U_c\rangle = \frac{1}{L\tau_0}\int_{-L/2}^{L/2}\int_{0}^{\tau_0} \textbf{r}_{t}\cdot \hat{\textbf{x}}\,ds\,dt.
\end{equation}
%%%
Here, forward swimming is when $\langle U_c\rangle <0$ and backward swimming is when $\langle U_c\rangle >0$. In Fig.~\ref{fig3}b, we show the contour plot of $|\langle U_c\rangle|$ as a function of $\eta$ and $\alpha_0$ with zero intrinsic curvature $\kappa_0 = 0$. Given an included angle $\alpha_0$, $\langle U_c \rangle$ is maximized around $\eta \approx 2.7$. At small $\eta$, the filaments are relatively rigid. The net translation over one period is small due to nearly reciprocal motions; at large $\eta$, viscous force dominates and the filament's deformation is confined around the actuation point (at $s=-L/2$) with the filament tail (at $s=L/2$) barely moving, leading to small propulsion. On the other hand, given $\eta$, there exists an optimal value of $\alpha_0$. The optimal $\alpha_0$ is around 1.1 when $\eta = 2.7$. When $\alpha_0$ approaches $\pi/2$ $(\approx 1.57)$, the propulsions from the two filaments nearly align with the $y$-axis and are opposite to each other. The cancellation between them leads to a small $\langle \textbf{U}_c \rangle$ along $x$ direction. At the opposite limit, where $\alpha_0 \to 0$, $\langle U_c \rangle$ is small due to small actuation amplitude.

The intrinsic curvature $\kappa_0$ has a strong effect on the swimming velocity (Fig.~\ref{fig4}a). When $\kappa_0<0$, the two filaments are curved inward at rest, and $|\langle U_c \rangle|$ is increased significantly. For $\kappa_0 = -1$, $|\langle U_c \rangle|$ is nearly tripled compared with $\kappa_0 = 0$. Figure~\ref{fig4}b shows that the beating pattern for $\kappa_0 = -1$ resembles that of cilia: compared with Fig.~\ref{fig3}a, the filaments are bent significantly during the recovery stroke, and the filaments stretch out straight during the power stroke. Over one period, the net displacement along the $-x$ direction is larger than that of $\kappa_0 = 0$. When $\kappa_0>0$, $|\langle U_c \rangle|$ decreases and the swimming direction is even reversed (backward swimming) for sufficiently large $\kappa_0$. The beating patterns shown in Fig.~\ref{fig4}c indicate that the power strokes become ineffective with larger deformation and thus yield less propulsion. But the recovery strokes become stronger, leading to a net displacement along $+x$ direction. The above effect of the intrinsic curvature is consistent with previous numerical simulations~\cite{Liu20}.
%%%%%%
\begin{figure}[t]
\centering
\includegraphics[bb=0 5 242 238, scale=1.0,draft=false]{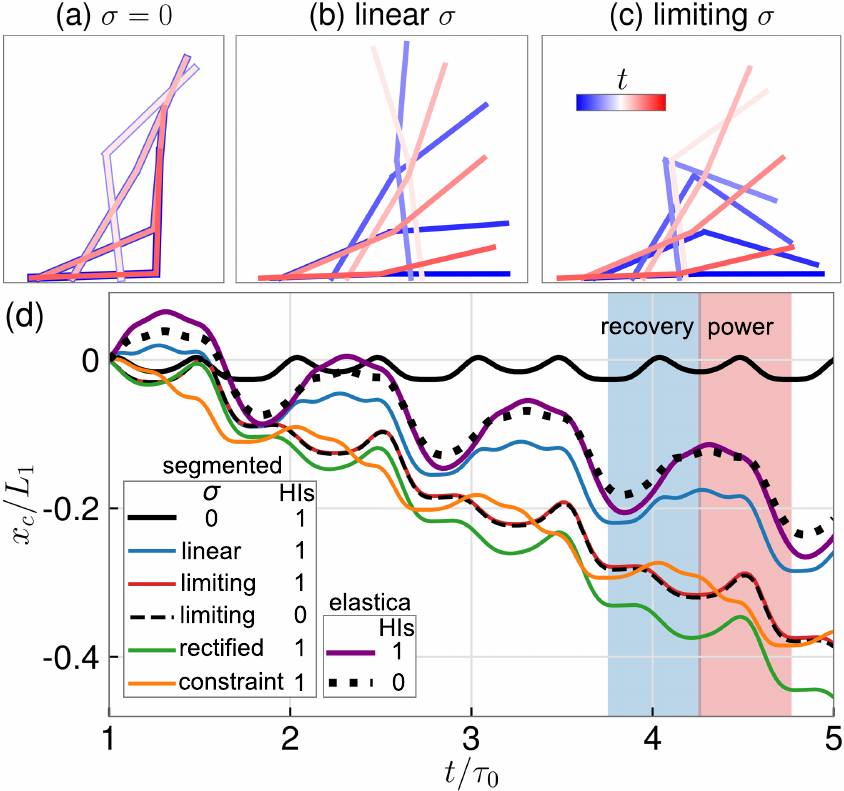}
\caption{Reciprocal and nonreciprocal motions of the clamshell swimmer with $\gamma = 0.5$, $\alpha_0 = 1.1$, and $c=10.0$. (a)--(c) Time lapse of the segmental model for (a) zero $\sigma$, (b) linear $\sigma$ with $\eta = 0.8$, and (c) limiting $\sigma$ with $\eta = 0.8$, $\sigma_n = -1.0$, and $\sigma_p = 0.15$. Blue: recovery stroke; red: power stroke. Time goes from blue to red. Only half swimmer is shown due to symmetry. In (a), recovery strokes overlap exactly with power strokes. See supplemental videos showing motions of the segmental model with different parameters. (d) The COM location $x_c/L_1$ as a function of time for the segmental model with different forms of torques including the three cases shown in (a)--(c), the rectified torque with $\eta = 0.8$ and $K/K'=15$, and the constraint torque with $\eta = 1.0$ and $\theta_m=0$. The elastic clamshell is also shown with $\eta=2.5$ and $\kappa_0 = 0$. `HIs = 1/0' corresponds to with and without hydrodynamic interactions between filaments. The initial time period is discarded with the positions at $t=1$ shifted to the origin.}
\label{fig5} 
\end{figure}
%%%%%

%%%%%%%%%%%%%%%%%%
\subsection{Segmental model}
%%%%%%%%%%%%%%%%%%

In the segmental model when the response torque at $J_2$ is zero, the motion is reciprocal and there is no net displacement over one period, as shown by the beating patterns in Fig.~\ref{fig5}a and the COM location $x_c(t)$ in Fig.~\ref{fig5}d (dark curve). With a linear torque, symmetric in both bending directions, the two-linked rigid segments resemble a flexible filament of zero intrinsic curvature. The beating patterns become non-reciprocal (Fig.~\ref{fig5}b) and the swimmer translates toward the $-x$ direction (blue curve). As a comparison, the displacement of the elastic clamshell with the velocity-optimal parameters is also shown (purple curve). Its backward displacement along the $+x$ direction during the recovery stroke is larger than that of the linear-torque swimmer. With the limiting torque and the rectified torque, the two-linked rigid segments resemble a flexible filament with nonzero intrinsic curvature. During the recovery stroke, the positive torque is limited by $\sigma_p$, which is smaller than the torque applied by the torsional spring when the relative deflection $\Delta \theta > \sigma_p/K$. This allows $\Delta \theta$ to reach larger values, and the orientation of segment 2 tends to align with its direction of translation, as shown in Fig.~\ref{fig5}c; therefore the viscous drag is reduced. As a result, the swimmer displacement after a full recovery stroke is smaller and even reversed toward the $-x$ direction, leading to a larger swimming speed. The beating patterns of the swimmer with either the rectified torque or constraint torque are similar to Fig.~\ref{fig5}c. Below we mainly focus on the results of the limiting torque.
%%%%%
\begin{figure}[t]
\centering
\includegraphics[bb=0 5 244 220, scale=1.0,draft=false]{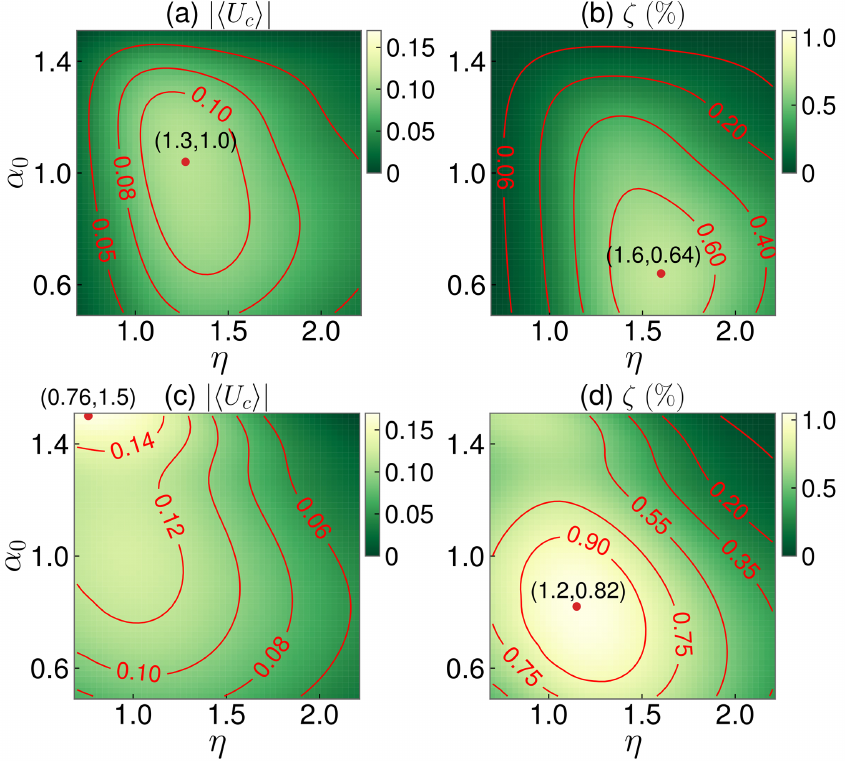}
\caption{Contour maps of $|\langle U_c \rangle|$ and $\zeta$ as functions of $\eta$ and $\alpha_0$ for the segmental swimmer with $\gamma = 0.5$ and $c=15.0$. (a), (b) The linear torque. (c), (d) The limiting torque.}
\label{fig6}
\end{figure}
%%%%%

The hydrodynamic interactions between filaments slightly enhance the swimming performance of the elastic clamshell, as shown by the difference between the dark dotted line and the purple line in Fig.~\ref{fig5}d. This is due to the fact that the velocity of each filament is opposite to the local induced velocity by other filament. When the clamshell opens, i.e., two filaments are moving away from each other, the HIs between them resist their separation. This induces additional deformations in the filaments that can reduce the viscous drag. However, in segmental model, the effect of HIs is negligible (dark dashed line), since the filaments are rigid and the torques due to HIs is small compared with the applied torques at $J_2$.

The time-averaged swimming speed of the segmental swimmer is given by $\langle U_c\rangle = \gamma\dot{x}_1^c+(1-\gamma)\dot{x}_2^c$, where $\dot{x}_1^c$ and $\dot{x}_2^c$ are the $x$-component COM velocities of the two segments. Similar to the definition in the previous work~\cite{Becker03}, we define the swimming efficiency as the ratio of the work needed to drag the swimmer with a fixed configuration at the average swimming speed $\langle U_c \rangle$ to the total work done by the displacements of the segments,
%%%
\begin{equation}\label{eq3.3}
\zeta = \frac{D \langle U_c \rangle}{\int_{-L_1/2}^{L_1/2}\textbf{f}_1 \cdot \dot{\textbf{r}}_1\,ds_1+\int_{-L_2/2}^{L_2/2}\textbf{f}_2 \cdot \dot{\textbf{r}}_2\,ds_2},
\end{equation}
%%%
where $D$ is the drag force experienced by the swimmer (without HIs) when translating with the initial configuration, i.e, $\theta_1 = \theta_2 = \alpha_0$, at the average speed $\langle U_c \rangle$. From Eq.~(\ref{eq2.3}), we obtain,
%%%
\begin{equation}\label{eq3.4}
D = 4\pi c^{-1}\langle U_c \rangle\left[1+(c-2)(c+2)^{-1}\sin^2\alpha_0\right].
\end{equation}
%%%
As shown by the contour maps in Figs.~\ref{fig6}a and \ref{fig6}b, for the linear-torque swimmer, both $|\langle U_c \rangle|$ and $\zeta$ have optimal values with respect to $\alpha_0$ and $\eta$. For $\gamma = 0.5$, the optimal efficiency is about $0.7 \%$ at $(\eta,\alpha_0) = (1.6, 0.64)$. For the swimmer with the limiting torque, the swimming performance is significantly improved. The swimming speed $|\langle U_c \rangle|$ has an optimal value in $\eta$ but increases as $\alpha_0$ is increased until the segments intersect each other. The optimal $\zeta$ is achieved at $(\eta,\alpha_0) = (1.2,0.82)$ and is more than $40\%$ larger than that of the linear-torque swimmer. 

%%%%%
\begin{figure}[t]
\centering
\includegraphics[bb=0 10 260 145, scale=0.85,draft=false]{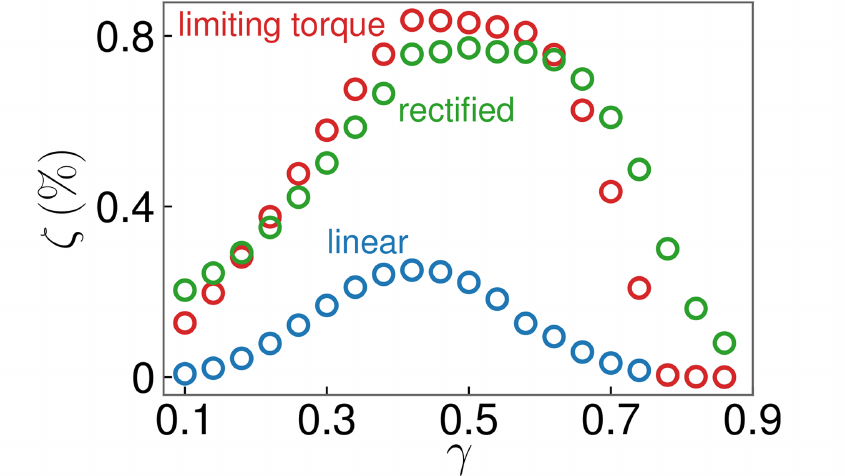}
\caption{Effect of the segment length ratio $\gamma$ on the swimming efficiency $\zeta$ with $\eta=1.0$ and $c=15.0$ for linear torque (blue symbols), limiting torque with $\sigma_n=-1.0$ and $\sigma_p=0.15$ (red symbols), and rectified torque with $K/K'=15.0$.}
\label{fig7}
\end{figure}
%%%%%
The segment length ratio $\gamma$ also has a strong effect on the swimming performance. As shown in Fig.~\ref{fig7}, optimal values of $\gamma$ exist, which is expected since the swimmer approaches a reciprocal scallop as $\gamma \to 0$ and $1$. For small $\gamma$, $L_1 < L_2$, and the amplitude of motion at $J_2$ is small due to small rotation radius around $J_1$. This is in analogy with the elastic clamshell of large $\eta$, in which the actuation is confined around $J_1$. For large $\gamma$, $L_1 > L_2$, the hydrodynamic torque upon segment 2 is small due to small segment length, and so does the response torque $\sigma (\theta_1, \theta_2)$, leading to small deflection angle $|\Delta \theta|$. This is in analogy with the elastic clamshell of small $\eta$ with small filament deformation. 

%%%%%
\begin{figure}[t]
\centering
\includegraphics[bb=0 5 245 192, scale=1.0,draft=false]{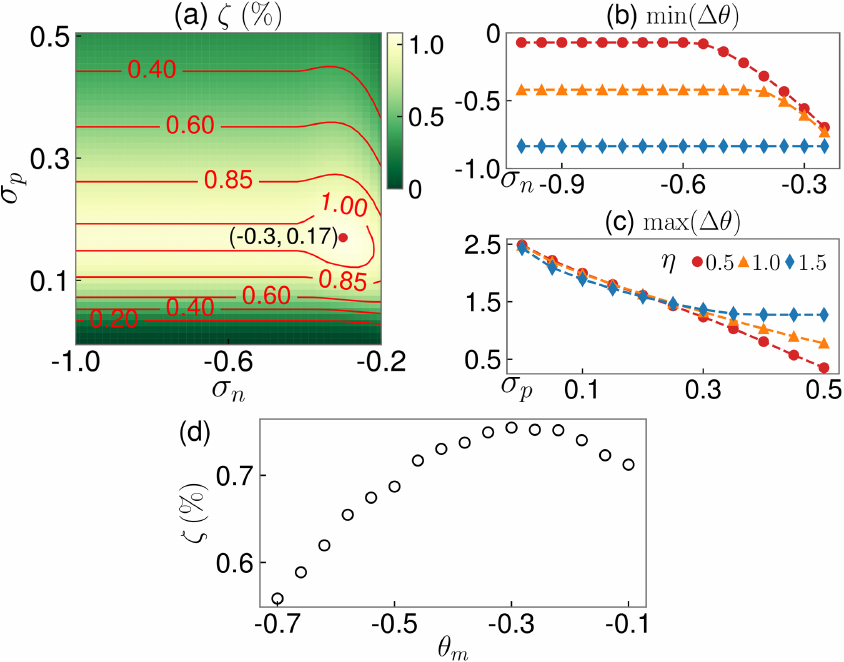}
\caption{(a) Contour map of $\zeta$ of the segmental model with the limiting torque as functions of $\sigma_n$ and $\sigma_p$ with $\gamma = 0.5$, $\alpha_0=0.8$, and $\eta = 1.0$. (b) $\min (\Delta \theta)$ as a function of $\sigma_n$ for different values of $\eta$ (see legend in (c)) with fixed $\sigma_p = 0.2$. (c) $\max (\Delta \theta)$ as a function of $\sigma_p$ for different values of $\eta$ with fixed $\sigma_n = -0.5$. (d) Effect of the threshold angle $\theta_m$ on $\zeta$ for the constraint torque with $\eta = 1.5$, $\gamma = 0.5$, and $\alpha_0 = 1.1$.}
\label{fig8}
\end{figure}
%%%%%
We now look at the effect of the torque-bias parameters, $\sigma_n$ and $\sigma_p$, on the swimming efficiency $\zeta$. As shown by the contour plot in Fig.~\ref{fig8}a, $\zeta$ has a maximum as a function of $\sigma_n$ and $\sigma_p$ at $(\sigma_n, \sigma_p) = (-0.3,\,0.17)$. During each period, $\sigma_n$ and $\sigma_p$ control the maximum and minimum deflections of segment 2 from segment 1, which can be measured by the maximum and minimum values of $\Delta \theta$ over one period. During the power stroke, as $\sigma_n$ decreases, the resistance for $\theta_2$ to be larger than $\theta_1$ increases; therefore $\min(\Delta \theta)$ increases (Fig.~\ref{fig8}b) and segment 2 becomes more aligned with segment 1. However, for sufficiently small $\sigma_n$, the hydrodynamic torque upon segment 2 may not reach $\sigma_n$ and $\min(\Delta \theta)$ becomes independent of $\sigma_n$. The existence of optimal values of $\sigma_n$ suggests optimal configurations for the power stroke. When $\eta = 1.0$ (orange triangles in Fig.~\ref{fig8}b), the optimal $\min(\Delta \theta) \approx -0.6$ at $\sigma_n = -0.3$, i.e., segment 2 is deflected about 34$^\circ$ counterclockwise relative to segment 1. Optimal power strokes are also observed for the segmental model with constraint torque (Fig.~\ref{fig8}d), as $\min(\Delta \theta)$ is controlled by the threshold angle $\theta_m$. During the recovery stroke, as $\sigma_p$ increases, the resistance for $\theta_2$ to be smaller than $\theta_1$ increases; therefore $\max(\Delta \theta)$ decreases (Fig.~\ref{fig8}c). Similar to the effect of $\sigma_n$, the existence of optimal values of $\sigma_p$ indicates optimal configurations for the recovery stroke. When $\eta=1.0$ (orange triangles in Fig.~\ref{fig8}c), the optimal $\max(\Delta \theta) \approx 1.7$ at $\sigma_p = 0.17$, i.e., segment 2 is deflected about 97$^{\circ}$ clockwise relative to segment 1, which is consistent with the observation from Fig.~\ref{fig5}c. Therefore, the controlled stroke patterns during power and recovery strokes by $\sigma_n$ and $\sigma_p$ are the key for the improved swimming performance. Figure~\ref{fig8}b and \ref{fig8}c also reveal strong dependence of $\min$/$\max(\Delta \theta)$ on $\eta$: for smaller values of $\eta$, $\min$/$\max(\Delta \theta)$ can vary appreciably by changing $\sigma_n$ or $\sigma_p$; but for sufficiently large $\eta$ (small $K$), they remain constant due to the domination of the linear part in $\sigma$ (Fig.~\ref{fig2}b). 

Finally, we report the sets of parameters that optimize the efficiency of the segmental model using the Nelder–Mead direct search method implemented in the SciPy \verb!optimize.minimize()! routine~\cite{Virtanen20}. For the limiting torque, the optimal parameters are $\gamma = 0.49(1)$, $\alpha_0 = 0.82(1)$, $\sigma_n = -0.29(1)$, $\sigma_p = 0.16(1)$, and $\eta = 0.99(1)$, with the optimized efficiency $\zeta = 1.10(1)\%$, where the small uncertainties on the second decimal place are due to different initial guesses and indicate the convergence to a global maximum. For the rectified torque, the optimal parameters are $\gamma = 0.48(0)$, $\alpha_0 = 0.80(1)$, $\eta = 1.10(1)$, and $K/K' = 5.17(2)$, with the optimized $\zeta = 1.22\%$.

%%%%%%%%%%%%%%%%%%%%%%
\section{Conclusions}\label{sec.4}
%%%%%%%%%%%%%%%%%%%%%%
In this work, we have numerically studied the dynamics of a two-arm clamshell swimmer at low Re with full hydrodynamic interactions, including an elastic clamshell constructed using flexible filaments and a segmental model constructed using rigid segments. Optimal elastoviscous numbers and included angles have been identified. In the segmental model, rigid segments are connected by mechanical joints with different response torques. The asymmetric beating patterns induced by the biased response torques significantly enhance the swimming performance. The effects of various parameters on the swimming efficiency have been extensively studied. Our results may be useful for the design and optimization of synthetic low-Re swimmers. 

The swimming performance may be further improved by optimizing the stroke pattern~\cite{Tam07}. The elastic clamshell may be optimized by considering varying stiffness along the filaments~\cite{Peng17}. Different from our coarse-grained model, a local curvature-dependent elastic modulus that is distributed along the arc length has been used in simulations to make the bending of a cilium easier towards one direction than the other and generate asymmetric beating patterns~\cite{Han18}. Minimum models have been constructed using rigid filaments and linear torsional springs for anchored boundary conditions to capture the main dynamics~\cite{Liu2_20,Zhu20,Canio17,Ling18}. A linear torsional spring has also been used to generate localized elasticity at the actuation point in an attempt to improve the propulsion of a single boundary-driven filament, but has been found to underperform compared to an elastic filament with distributed elasticity~\cite{Peng17}. The limbs of many crustaceans, like shrimp and crayfish, consist of linked rigid segments and beat asymmetrically as our segmental model. To mimic the asymmetric beating pattern, rigid paddles are treated as impermeable during the power stroke and permeable during the recovery stroke in simulations~\cite{Zhang14}. The effect of hydrodynamic interactions on the dynamics of the segmental model has been shown to be negligible, but may become important as the number of arms increases and the separations between them decrease.

\section*{Conflicts of interest}
There are no conflicts of interest to declare.

\section*{Acknowledgements}
We thank Leif Ristroph and Stephen Childress for inspiring questions and helpful discussions. We thank Yilin Li for demonstration experiments. S.H. gratefully acknowledges support from the MacCracken Fellowship and the Global Research Initiatives Fellowship provided by New York University. We also thank the support from NYU-ECNU research institute at NYU Shanghai. MJS acknowledges support by the National Science Foundation under awards DMR-1420073 (NYU MRSEC) and DMR-2004469.

%%%%%%%%%%%%%%%%%%%%%%%%%%%%%%%%%%%%%%%%%%%%%%%%%%%%%%
\appendix
\renewcommand{\theequation}{\thesection\arabic{equation}}
\renewcommand{\thefigure}{\thesection\arabic{figure}}
%%%%%%%%%%%%%%%%%%%%%%%%%%%%%%%%%%%%%%%%%%%%%%%%%%%%%%
\newpage
\section{Boundary conditions}\label{sec.a}
\setcounter{equation}{0}
%%%%%%%%%%%%%%%%%%%%%%%%%%%%%%%%%%%%%%%%%%%%%%%%%%%%%%
First, at $s=-L/2$, we have,
%%%%
\begin{equation}\label{Eq.a1}
\theta(s=-L/2) = \alpha(t),
\end{equation}
%%%%
Take a variational derivative of the filament energy with respect to an arbitrary shape deformation $\delta \textbf{r}$,
%%%%%
\begin{equation}\label{Eq.a2}
\begin{aligned}
\delta \mathcal{E} = \,&B(\kappa \textbf{p}^{\perp}-\kappa_0\textbf{p}^{\perp})\cdot \delta \textbf{r}_{s}\big|^{L/2}_{-L/2}\\
&-\left(B\textbf{r}_{sss}+B \kappa_0 \kappa\textbf{p}-T\textbf{p}\right)\cdot \delta \textbf{r}\big|_{-L/2}^{L/2}\\
&+\int_{-L/2}^{L/2} \left[B\textbf{r}_{ssss}+B\kappa_0\left(\kappa\textbf{p}\right)_{s}-\left(T\textbf{p}\right)_{s}\right]\cdot \delta \textbf{r}\,ds.
\end{aligned}
\end{equation}
%%%%
With no constraints, the boundary conditions at $s=L/2$ can be obtained from the first two terms on the r.h.s. of Eq.~(\ref{Eq.a2}),
%%%%
\begin{equation}\label{Eq.a3}
\theta_s = \kappa_0,\ \theta_{ss} = 0,\text{ and } T = 0,\text{ at }s=L/2.
\end{equation}
At $s=-L/2$, the $y$-component of the filament force is cancelled due to the mirror symmetry, so we require the $x$-component force to be zero,
%%%%
\begin{equation}\label{Eq.a4}
2\left(-B\textbf{r}_{sss}-B\kappa_0\kappa\textbf{p}+T\textbf{p}\right)\cdot \hat{\textbf{x}} = 0,
\end{equation}
%%%%
which can be interpreted as a boundary condition for $T$,
%%%%
\begin{equation}\label{Eq.a5}
T = -B\theta_{ss}\tan \theta-B\theta_{s}^2+B\kappa_0\theta_{s},\ \text{at }s=-L/2.
\end{equation}
%%%%
To keep the separation of the two filaments fixed, we enforce the $y$-component velocity of the filament at $s=-L/2$ to be zero, $y_{t}(s=-L/2) = 0$, which generates a boundary condition for $\theta$,
%%%%
\begin{equation}\label{Eq.a6}
\begin{aligned}
\theta_{sss} =\, &(1+2c^{-1})^{-1}\Big[(5\theta_{s}\theta_{ss}-2\kappa_0 \theta_{ss}-2c^{-1}\theta_s\theta_{ss}\\
&+2B^{-1} T_{s})\tan \theta+\frac{8\pi\mu B^{-1} c^{-1}\textbf{U}\cdot\hat{\textbf{y}}}{
\cos\theta}\Big].
\end{aligned}
\end{equation}
%%%%

%%%%%%%%%%%%%%%%%%%%%%%%%%%%%%%%%%%%%%%%%%%%%%%%%%%%%%%
\section{Numerical methods}\label{sec.b}
\setcounter{equation}{0}
%%%%%%%%%%%%%%%%%%%%%%%%%%%%%%%%%%%%%%%%%%%%%%%%%%%%%%%
We solve the system of governing equations using a finite difference method. Discretize the arc length with a uniform grid, $s_j = j/N$-1/2 with $j = 0$, 1, $\cdots$, $N$, and denote the quantities at $s_j$ with subscript $j$. The spatial derivatives are approximated using a second-order scheme. We discretize time as $t_n = n\Delta t$ and denote with superscript $n$ the quantities at the current time step $t_n$. Given the filament position $\textbf{r}^{n}$, filament tension $T^{n}$, and $\theta^{n+1}_0 = \alpha^{n+1}$, we solve for $\theta^{n+1}$ and $T^{n+1}$. The $\theta$ equation [Eq.~(\ref{eq2.5})] is a fourth order partial differential equation with a nonlinear boundary condition [Eq.~(\ref{Eq.a6})]. To avoid the strict fourth-order stability limit for the time-step size, we treat $\theta_{ssss}$ implicitly and use a second-order backward differentiation formula for the time stepping. The remaining terms such as lower order derivatives, tension, and the nonlocal integrals [Eqs.~(\ref{eq2.7}) and (\ref{eq2.8})] are extrapolated from previous time steps. Schematically, we write, 
\begin{equation}\label{Eq.b1}
\theta^{n+1}+\beta \theta^{n+1}_{ssss} = p^{n,n-1},
\end{equation}
where $\beta$ is a constant depending on $\Delta t$. We then split $\theta$ into two terms,
\begin{gather}
\theta^{n+1} = (\theta^P)^{n+1}+(\theta_{sss})_0^{n+1}\theta^{H},\ \text{with} \label{Eq.b2}\\
(\theta^P)^{n+1} + \beta (\theta^P_{ssss})^{n+1} = p^{n,n-1}, \label{Eq.b3}\ \text{and}\\
\theta^H + \beta\theta^H_{ssss} = 0. \label{Eq.b4}
\end{gather}
The boundary conditions of $\theta^{P}$ and $\theta^{H}$ can be inferred from the boundary conditions of $\theta$ and are linear. With $(\theta^P)^{n+1}$ and $\theta^H$ (only need to form once), we can form $\theta^{n+1}$ if $(\theta_{sss})_0^{n+1}$ is known. Since both $(\theta_{sss})_0$ and $T_{0}$ are nonlinear functions of $(\theta_{s})_0$ and $(\theta_{ss})_0$, we solve the tension equation [Eq.~(\ref{eq2.6})] together with Eqs.~(\ref{Eq.a5}) and (\ref{Eq.a6}) for $T^{n+1}$, $(\theta_{s})_0^{n+1}$, and $(\theta_{ss})_0^{n+1}$ using Newton's method. Let superscript $k$ denote current solutions at the $k$-th Newton's iteration. We linearize the tension equation and obtain a system of linear equations for the update $\delta T$,
\begin{gather}\label{Eq.b5}
\delta T_{ss}+M^{k}\delta T = Q^{k},\ \text{with}\\
\delta (T_{s})_0+m^{k} \delta T_0 = q^{k}\ \text{and}\ \delta T_{N} = 0,
\end{gather}
where $M^{k}$, $Q^{k}$, $m^{k}$, $q^{k}$ are functions of the current solutions. Solving for $\delta T$, updating $T_k$, $(\theta_{s})_0^{k}$, and $(\theta_{ss})_0^{k}$, and iterating until converge, we obtain $T^{n+1}$ and $\theta^{n+1}$. The above numerical scheme is second-order accurate both in space and time. For most of our simulations, we use $N=101$ and $\Delta t=10^{-4}$--$10^{-2}$.

%%%END OF MAIN TEXT%%%


\balance


%%%REFERENCES%%%
\bibliography{rsc} %You need to replace "rsc" on this line with the name of your .bib file
\bibliographystyle{rsc} %the RSC's .bst file

\end{document}